\documentclass[
    ,final            
  ]
  {aipproc}

\layoutstyle{8x11single}

\begin{document}

\title{Broken Valence Chiral Symmetry and Chiral Polarization of 
Dirac Spectrum in N$_f$=12 QCD at Small Quark Mass}


\author{Andrei Alexandru}{
  address={George Washington University, Washington, DC, USA}
}

\author{Ivan Horv\'ath}{
  address={University of Kentucky, Lexington, KY, USA [the speaker]}
}

\begin{abstract}
   The validity of recently proposed equivalence between valence spontaneous chiral symmetry 
   breaking 
   (vSChSB) and chiral polarization of low energy Dirac spectrum (ChP) in SU(3) gauge theory, 
   is examined for the case of twelve mass--degenerate fundamental quark flavors. We find 
   that the vSChSB--ChP correspondence holds for regularized systems studied. Moreover, our 
   results suggest that vSChSB occurs in two qualitatively different circumstances: there 
   is a quark mass $m_c$ such that for $m > m_c$ the mode condensing Dirac spectrum exhibits 
   standard monotonically increasing density, while for $m_{ch} < m < m_c$ the peak around 
   zero separates from the bulk of the spectrum, with density showing a pronounced depletion 
   at intermediate scales. Valence chiral symmetry restoration may occur at yet smaller 
   masses $m < m_{ch}$, but this has not yet been seen by overlap valence probe, leaving 
   the $m_{ch}=0$ possibility open. The latter option could place massless N$_f$=12 theory 
   outside of conformal window. Anomalous behavior of overlap Dirac spectrum for 
   $m_{ch} < m < m_c$ is qualitatively similar to one observed previously in zero and few--flavor 
   theories as an effect of thermal agitation.

\end{abstract}

\maketitle


\section{Introduction}

Vacuum effects play an important role in physics of strong interactions. Indeed, 
confinement and spontaneous chiral symmetry breaking (SChSB) are vital for current 
understanding of hadronic physics. However, attempts to clarify the origin and dynamics 
of these basic phenomena have not yet produced desirable results. Lattice QCD 
is expected to play a decisive role in achieving robust progress since the relevant 
information, stored in statistical ensembles of regularized systems, is generated as 
a byproduct of any numerical simulation. Extracting and interpreting this information 
in an unbiased  model--independent way underlies the ``bottom--up'' strategy for studying 
QCD vacuum structure~\cite{Hor06B}.

Focusing on chiral symmetry breaking, we follow the {\em valence} approach, wherein 
the gauge vacuum of arbitrarily massive theory is probed (by external massless quarks
propagating on its backgrounds) for long--range correlations necessary to break
valence chiral symmetry. This is reasonable~\cite{Ale12D,Ale14A}, since the symmetry
interpretation is maintained in completely general case, and valence SChSB (vSChSB) turns 
into standard SChSB when at least two physical quarks are massless. The presence or 
absence of vSChSB distinguishes qualitatively different states of vacuum in non--Abelian 
gauge theories: the ``real world'' QCD is strongly believed to be broken in this sense, 
and the dynamics of other SU(3) theories is guaranteed to be significantly different 
if they happen to be symmetric. 

Performing manipulations leading to standard Banks--Casher relation~\cite{Ban80A},
one can easily inspect that the connection of SChSB to Dirac mode condensation
naturally extends to the valence case. Indeed, valence chiral condensate appears 
if and only if mode condensate does, i.e. if spectral density of modes tends 
to non--zero value in the vicinity of the origin. While being only kinematic, 
i.e. true independently of underlying dynamical mechanisms, Banks--Casher equivalence 
provides for a useful tool to detect vSChSB. However, a different equivalence of 
similar scope has recently been suggested~\cite{Ale12D,Ale14A}, that is dynamical 
in the sense that its validity depends on specific realization of vSChSB in SU(3) gauge 
theories. This {\em vSChSB-ChP correspondence} proposes that chiral symmetry is broken 
if and only if the low end of Dirac spectrum exhibits chiral polarization (ChP) 
of the dynamical type described in Ref.~\cite{Ale10A}.

Potential validity of the above connection offers valuable insight into chiral symmetry
breaking phenomenon. Indeed, among other things, it formalizes and validates 
the intuitively appealing notion that vSChSB is the same effect as condensation of 
chirality~\cite{Ale12D}. It also places non--trivial constraints on possible 
explanations of SChSB which would have to be consistent with it to become viable. 
We emphasize that, in the context of SU(3) theories, vSChSB--ChP correspondence is 
expected to hold for arbitrary number and masses of fundamental flavors. Such broad 
association underlies the fact that SChSB mechanism should exhibit certain universality 
within the same class of interactions. 

At this time, vSChSB--ChP correspondence has been confirmed to hold in N$_f$=0 theory 
both at zero and finite temperature~\cite{Ale12D,Ale14A}, in zero--temperature N$_f$=2+1 
close to physical point~\cite{Ale12D}, and also in zero--temperature N$_f$=12 down
to fairly small quark masses~\cite{Ale14A}. In this presentation, we focus on the last 
case which is not only important in the context of vSChSB--ChP correspondence, but 
also in its own right. Indeed, massless N$_f$=12 theory remains controversial as 
a possible borderline case of dynamics within conformal 
window~\cite{Nei12A,Gie12A,Kut13A}. However, our results suggest a rather unexpected 
``phase structure'' in the behavior of Dirac spectrum as the mass is lowered toward 
chiral limit. These findings may have non--trivial implications regarding t
he possibility of low--energy conformality in this system.

\section{Dirac Mode Condensation and Chiral Polarization}

The results presented here derive from two cumulative spectral densities, namely 
\begin{equation}
  \sigma(\lambda,V) \equiv \frac{1}{V} \, 
  \langle \, \sum_{0 < \lambda_k < \lambda} \, 1\;\rangle
  \qquad\qquad
  \sigma_{ch}(\lambda,V) \equiv \frac{1}{V} \, 
    \langle \, \sum_{0 < \lambda_k < \lambda} \, C_{A,k}\;\rangle
  \label{eq:cumulative}
\end{equation}
where $\lambda \ge 0$ and $\lambda_k$ (real numbers) are imaginary parts of Dirac 
eigenvalues or, as frequently done in lattice studies with overlap Dirac operator, 
magnitudes of the eigenvalues weighted by the sign of imaginary part. Note that 
$\sigma$ is the standard cumulative mode density of Dirac operator while $\sigma_{ch}$ 
requires some explanation.

The object $C_{A,k}$ associated with every eigenmode $\psi_k$ is its correlation 
coefficient of polarization~\cite{Ale10A}. It expresses the tendency for asymmetry
(polarization) in local magnitudes of left--right components, relative to statistically 
independent pairing. To explain its construction, consider the process of selecting 
a sample from the former (correlated) and the latter (uncorrelated) distributions
and comparing their polarizations. Statistics of such comparisons determines 
the probability $\Gamma_{A,k}$ that sample from correlated distribution is more 
polarized than sample from uncorrelated one. The correlation coefficient of polarization
in mode $\psi_k$ is then $C_{A,k} = 2 \Gamma_{A,k}-1$. Thus, $C_{A,k}>0$ signifies 
enhancement of local chirality relative to statistical independence (dynamical 
chirality), while $C_{A,k}<0$ implies its suppression (dynamical anti--chirality).
An important virtue of this definition is that it does not depend on how we
choose to quantify the polarization of samples (reparametrization invariance).

Low--energy Dirac spectrum is chirally polarized when low--lying Dirac modes tend
to be positively correlated in the above sense. In such case, the cumulative chirality 
function $\sigma_{ch}(\lambda)$ starts at $\lambda=0$ as monotonically increasing, 
growing until it reaches the chiral polarization scale at $\lambda=\Lambda_{ch}$ 
where anti--correlation sets in. This creates a positive ``bump'' in 
$\sigma_{ch}(\lambda)$ with maximum value $\Omega_{ch}$ achieved at $\Lambda_{ch}$. 
The quantity $\Omega \equiv \sigma(\Lambda_{ch})$ represents the total volume density 
of chirally polarized modes. Note that, in finite volume, the three non--negative 
characteristics $\Lambda_{ch}$, $\Omega_{ch}$ and $\Omega$ can only be positive 
simultaneously. They thus serve as {\em finite--volume order parameters} of chiral 
polarization and, via most generic form of vSChSB--ChP correspondence 
({\em Conjecture 3} of Ref.~\cite{Ale14A}), as finite--volume order parameters 
of chiral symmetry breaking. It is frequently useful to eliminate $\lambda$ from 
relations \eqref{eq:cumulative}, and consider $\sigma_{ch}=\sigma_{ch}(\sigma)$. 
Utilizing this dependence is more efficient in situations involving depleted regions 
of spectrum, while ChP is still being signaled by a characteristic positive bump. 
Note that $\Omega$ and $\Omega_{ch}$ can be determined directly from 
$\sigma_{ch}(\sigma)$.

Spectral mode density $\rho(\lambda,V)$ is the right derivative of $\sigma(\lambda,V)$
with respect to $\lambda$. Coarse--graining in $\lambda$ is necessary in practice and, 
in absence of suspicion for singularity, we use the symmetric discretization, i.e.
\begin{equation}
    \rho(\lambda,\delta\lambda,V) \equiv 
    \frac{\sigma(\lambda+\delta\lambda/2,V) - \sigma(\lambda-\delta\lambda/2,V)}
         {\delta\lambda}  \qquad\quad
         \lambda \ge \delta\lambda/2
    \label{eq:rho}
\end{equation}
In terms of the above, Dirac mode condensation in arbitrary theory occurs when
\begin{equation}
   \lim_{\lambda \to 0} \, \lim_{\delta\lambda \to 0} \, \lim_{V\to \infty} \,
   \rho(\lambda,\delta\lambda,V) > 0
   \label{eq:condensation}
\end{equation} 
i.e. when abundance of ``infinitely infrared'' modes scales the same way as total 
number of modes, namely with volume. Using the Banks--Casher equivalence, the behavior
of $\rho(\lambda,V)$ in the vicinity of $\lambda=0$ at sufficiently large volumes
can be used to infer the presence or absence of vSChSB. 

Note that exact zeromodes are not included in spectral sums of Eq.~\eqref{eq:cumulative}.
Indeed, representing an infinite--volume quantity, chiral condensate will not be influenced 
by their omission. Moreover, zeromodes are chiral {\em a priori}, and their chirality is 
thus not an acceptable dynamical indicator of chirally polarized dynamics. One should thus 
keep in mind that vSChSB--ChP correspondence is concerned with the behavior of non--zero 
modes: ``infinitely infrared'' ones in case of vSChSB, and low--lying ones (band of width 
$\Lambda_{ch}$) in case of ChP.

\section{Lattice Technicalities}

Being interested in chirality effects, it is prudent to define probing valence
quarks via lattice Dirac operator with good chiral properties. Overlap Dirac 
operator~\cite{Neu98BA} is thus optimal for this purpose, and was used in this study. 
Taking into account that many light flavors tend to make gauge fields rather rough 
in current simulations, the appropriately large value of $\rho$--parameter (1.55) was 
set in the overlap construction based on standard Wilson--Dirac operator. We used 
implicitly restarted Arnoldi method with deflation to calculate lowest 200 modes with 
non--negative imaginary parts for each configuration in ensembles described below. 

To carry this study out, we obtained a subset of staggered fermion ensembles described 
in Ref.~\cite{Has12B}. Apart from fundamental plaquette with coupling constant $\beta_F$, 
the gauge action used also involves a negative adjoint plaquette term coupled via $\beta_A$. 
Among other things, this arrangement helps avoid the appearance of phases 
generated by lattice artifacts. As studied by the authors of Ref.~\cite{Has12B}, 
lattice theories at couplings described in Table~\ref{tab:stag_ensemb} are free of such 
spurious effects. Standard staggered fermion action with nHYP smearing was used in 
the dynamical quark sector. Configurations were identically smeared also in conjunction 
with our overlap valence quark calculations.  

Using the quark--mass scans in Ref.~\cite{Has12B}, we selected the representative 
of the ``heavy mass'', at which spectral density of staggered operator was observed
to have the standard low--flavor behavior with apparent mode condensation, and thus
vSChSB. This is the ensemble $S_1$ of Table~\ref{tab:stag_ensemb} . In the conformal 
scenario for massless N$_f$=12, it is expected that there is a non--zero mass $m_{ch}$
such that for $m<m_{ch}$, valence chiral symmetry becomes restored. This was indeed
observed in Ref.~\cite{Has12B} at the ``light mass'' of $am=0.0025$, where the behavior
of staggered spectral density is consistent with absence of mode condensation. 
The corresponding ensembles are $S_2$ and $S_3$ of Table~\ref{tab:stag_ensemb}, 
representing the system at two different volumes. 

\begin{table}[b]
   \centering
   \begin{tabular}{@{} cccccccccc @{}} 
      \hline
      Ensemble  &  Size  &  $\beta_F$  &  $\beta_A/\beta_F$  &  $a m$  &  N$_{conf}$  &
      $a \lambda_{min}^{av}$  &  $a \lambda_{min}$ &  
      $a \lambda_{max}^{av}$  &  $a \lambda_{max}$\\
      \hline
     $S_{1}$  &  $16^{3}\times 32$  & 2.8  & -0.25  &  0.0200  &  100  &  
                                      0.0098  &  0.0007  & 0.5190  &  0.5259\\
     $S_{2}$  &  $32^{3}\times 64$  & 2.8  & -0.25  &  0.0025  &  30   &  
                                      0.0007  &  0.0001  &  0.2016  &  0.2035\\
     $S_{3}$  &  $24^{3}\times 48$  & 2.8  & -0.25  &  0.0025  &  50   &
                                      0.0193  &  0.0001  &  0.3046  &  0.3075\\  
      \hline
   \end{tabular}
   \caption{Ensembles of N$_f$=12 lattice QCD with description in the text. 
   Note that $\lambda_{min}^{av}$ denotes the average lowest magnitude of eigenvalue in 
   a configuration, while $\lambda_{min}$ is the overall minimum in a given ensemble. 
   Similarly for the maximal eigenvalues computed.}
   \label{tab:stag_ensemb}
\end{table}

\section{Results}

\begin{figure}
  \includegraphics[height=.5\textheight]{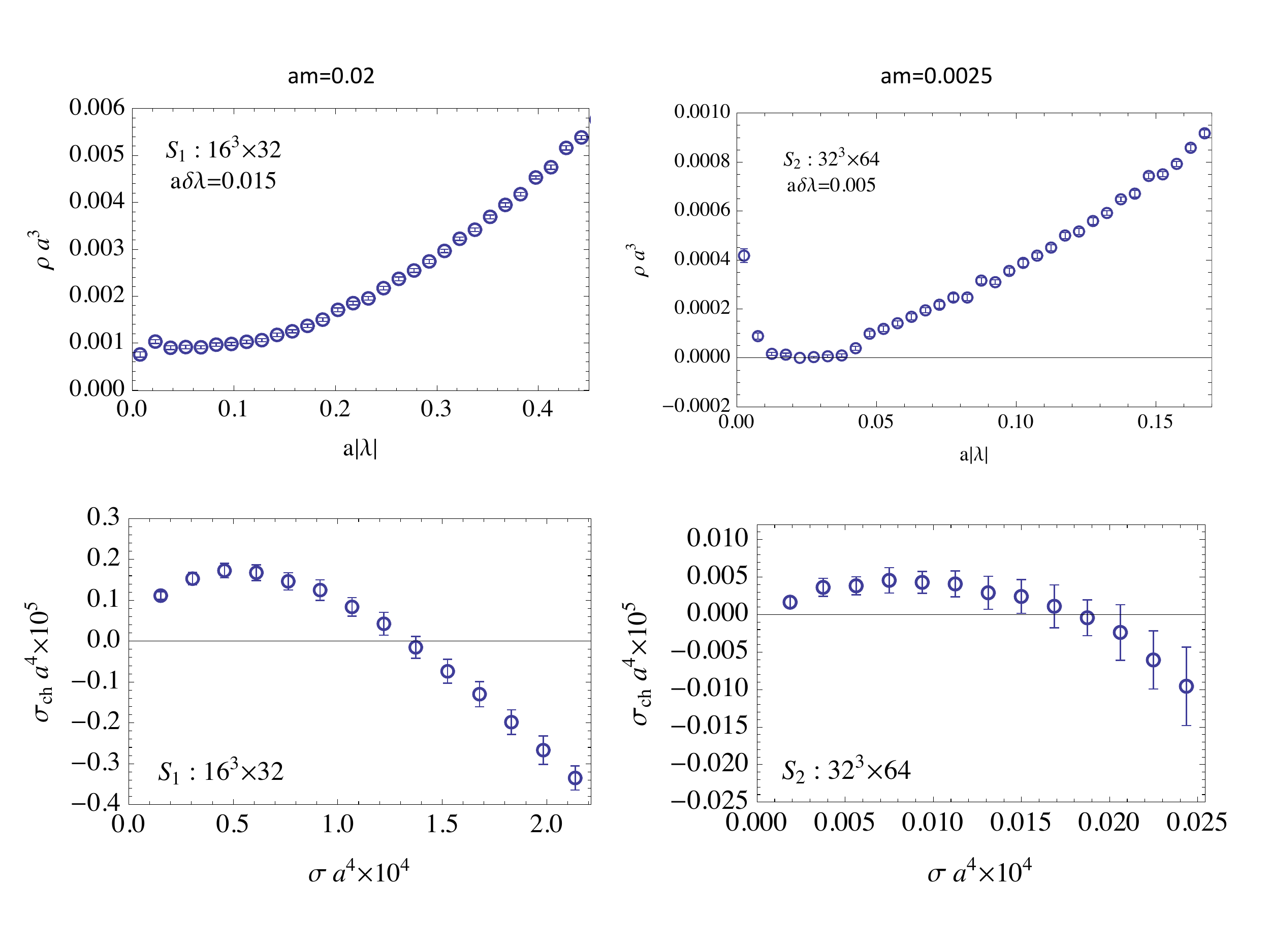}
  \label{fig:first} 
  \caption{Comparison of spectral density and chiral polarization in ensembles
           $S_1$ (``heavy'', left) and $S_2$ (``light'', right).}
\end{figure}

The relevant results for ``heavy'' ensemble ($S_1)$ are shown in 
Fig.~\ref{fig:first} (left). Similarly to staggered spectrum of Ref.~\cite{Has12B},
we find the behavior in overlap spectral density typical of mode--condensing theory, 
with clear tendency for non--zero intercept and the standard overall shape. There is 
thus little doubt that we are dealing with chirally broken case. At the same time,
the cumulative chirality $\sigma_{ch}(\sigma)$ exhibits the positive ``bump'' 
(bottom left), revealing that the theory is chirally polarized. Thus, vSChSB--ChP 
correspondence holds in this case.

Analogous results for the ``light'' ensemble at larger of the two volumes ($S_2$) 
are displayed in Fig.~\ref{fig:first} (right). The surprise here is that instead
of spectral density falling toward zero near origin, expected from staggered quark
study~\cite{Has12B}, overlap spectrum clearly displays mode--condensing behavior. 
Thus, at least in this volume, the vacuum of the theory looks chirally symmetric 
when probed by staggered valence quarks, but it responds as clearly broken when 
overlap valence quarks are used. Nevertheless, with that being the case, vSChSB--ChP 
correspondence demands that overlap Dirac spectrum be chirally polarized. As one can 
see in Fig.~\ref{fig:first} (bottom right), this is indeed the case. Note that 
the total polarization has dropped significantly relative to the ``heavy'' case but 
this is not surprising given the overall reduction in mode abundance at low energy.

An important part of the above finding is the highly non--standard overall shape
of overlap spectral density in the light quark regime. The separation of the spectrum
into two parts, namely the very infrared peak with spectral density falling 
away from zero and the monotonically increasing bulk receded to higher energies,
is in stark contrast to typical chirally broken spectra such as that of $S_1$.   
We refer to the dynamics exhibiting globally non--monotonic behavior in spectral 
density of the above type, as well as the density itself, as {\em anomalous}. 
An intriguing point in this regard is that the spectral density of this kind was first 
observed some time ago in a very different physical context: in N$_f$=0 theory at 
finite temperature past the deconfinement temperature $T_c$~\cite{Edw99A}. 

\begin{figure}
  \includegraphics[height=.5\textheight]{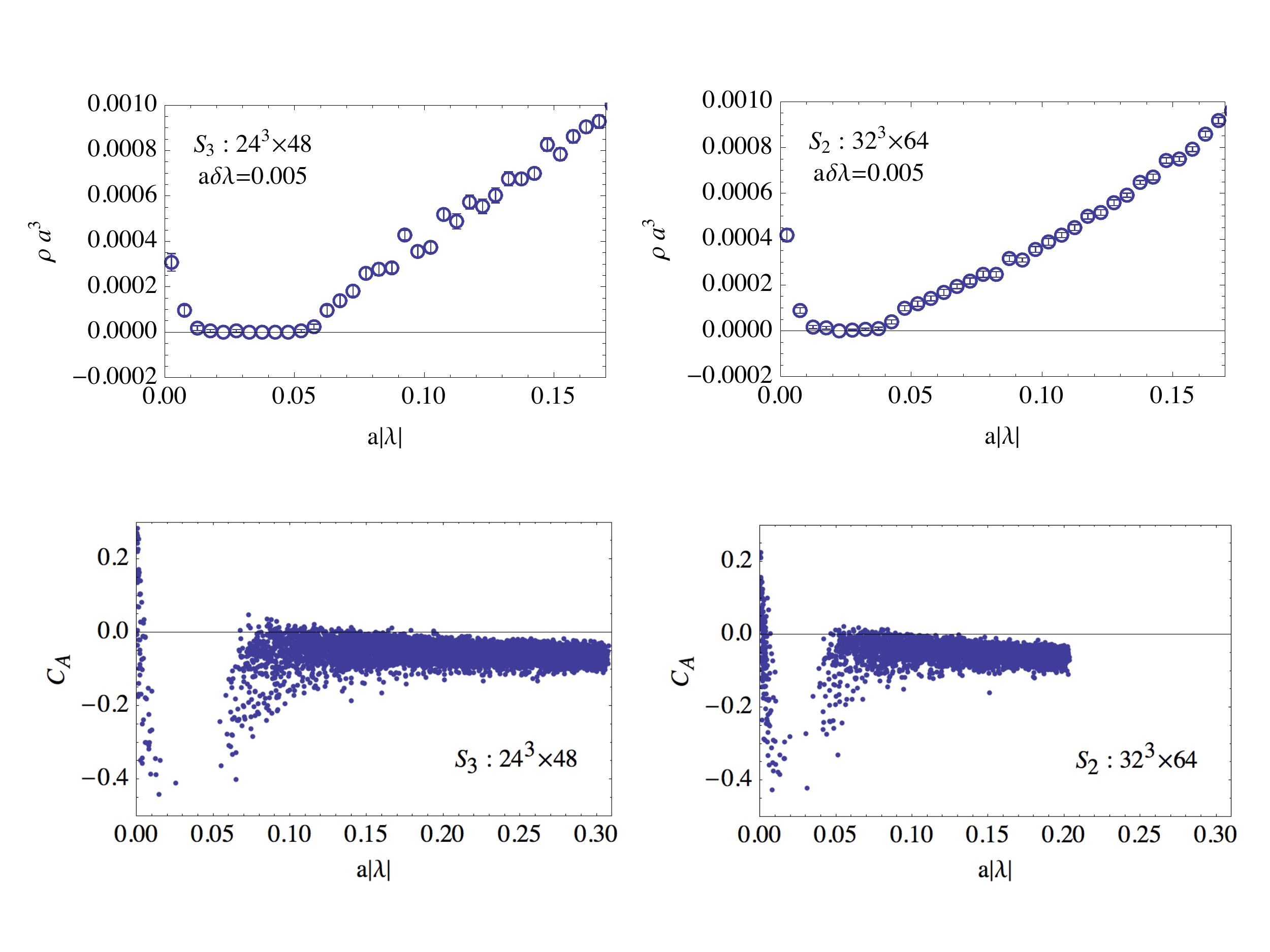}
  \label{fig:second} 
  \caption{Volume comparison for spectral densities and scatter plots of 
           chiral polarization with ensembles $S_3$ and $S_2$.}
\end{figure}

The observed anomalous behavior at light mass could in principle result from infrared 
cutoff being too high, i.e. to be a finite volume effect. In the staggered case, 
the change from $24^3\times 48$ lattice to $32^3\times 64$ lattice ($S_3$ to $S_2$) 
produced little change in the behavior of spectral density~\cite{Has12B}. 
Fig.~\ref{fig:second} (top) shows this comparison in the overlap case. The most 
important aspect to check in this regard is whether there is a large reduction 
in the anomalous peak of near--zeromodes, which could suggest the possibility
of an infrared artifact. However, as the plots reveal, this does not happen. 
In fact, the height of the peak increases in larger volume.   

Note that in the lower part of Fig.~\ref{fig:second} we show chiral polarization 
effects as scatter plots in $\lambda$--$C_A$ plane. Here each dot corresponds to 
one eigenmode, and all eigenmodes computed for these ensembles are included. 
The chirally polarized infrared peak, the dramatic break in the spectrum at 
intermediate scales (eigenmode depletion), and the anti--polarized bulk are 
seen here clearly and from somewhat different perspective. 

\section{Discussion}

In this study of 12--flavor SU(3) theory we set out to check whether 
dynamical effects of many light flavors are consistent with
the proposition that valence chiral symmetry breaking and chiral 
polarization are always locked together in the realm of SU(3)
theories with fundamental quarks: the vSChSB--ChP correspondence.
Our conclusion from this pilot study~\cite{Ale14A} is that this is
indeed the case, thus strengthening the merits of the conjecture. 

However, the results obtained for overlap Dirac spectra reveal considerably 
more about SU(3) dynamics, possibly with far--reaching consequences. Indeed, 
while looking for transition from broken to symmetric theory when crossing 
from ``heavy'' to ``light'', we rather found the transition from standard 
to anomalous behavior of spectral density. Even though vSChSB apparently 
remains in the light regime (at least for some band of light masses), 
a drastic transition nevertheless occurs: the vacuum qualitatively 
changes and the dynamics of propagating fermions does as well. Given 
that, we predict the existence of mass $m_c>0$ associated with this 
transition, and expect that, among other things, the nature of vSChSB 
qualitatively changes as the mass is lowered below $m_c$.
\footnote{Note that in Ref.~\cite{Ale14A}, the mass $m_c$ was denoted 
as $m_{in}$.}

While unlikely an infrared artifact as we checked, one can legitimately 
ask whether the anomalous behavior could be attributed to ultraviolet 
mutilation, albeit the other known types of lattice effects were 
avoided by the current choice of couplings. Although this will certainly 
be studied further, it is relevant in this regard that very similar 
anomalous phase appears in N$_f$=0 system subjected to thermal agitation, 
in some temperature band beyond $T_c$ \cite{Edw99A,Ale12D,Ale14A}.
That occurrence has recently been studied in detail, producing a rather
convincing evidence that the anomalous phase is neither an infrared 
(Ref.~\cite{Ale14A,Ale14D}) nor an ultraviolet (Refs.~\cite{Ale14B, Ale14D})
artifact, and is thus most likely physical. This makes it very reasonable 
to interpret our finding as a manifestation of the fact that many light 
quarks and thermal bath have very similar disordering effects on chirality, 
and that we have simply encountered another corner of the theory space where 
anomalous phase occurs.

Given the expectation that the anomalous phase $m_{ch}<m<m_c$ can indeed 
materialize in continuum N$_f$=12 system, it is of great interest to 
understand the nature of its dynamics. Note that in the N$_f$=0
thermal case, where both vSChSB and confinement have well--defined symmetry
interpretation, the most basic characterization of its anomalous regime is 
that it is a chirally broken phase without confinement. While (de)confinement 
is not straightforward to infer on its own in the light 12--flavor case, 
it is natural to expect that, at the very least, the processes related 
to deconfinement take place as the quark mass is lowered below $m_c$ 
in that system.

An intriguing question we are obviously left with is whether valence chiral 
symmetry ever gets restored in N$_f$=12. In other words, whether there 
is a second, yet lighter regime $0<m<m_{ch}>0$, where anomalous Dirac spectrum 
disappears and there is no mode condensation. In thermal N$_f$=0 case it is 
reasonable to adopt the view that there is finite temperature $T_{ch}>T_c$ 
beyond which this happens. Indeed, the associated lack of near--zeromode 
accumulation has been confirmed with overlap operator at sufficiently large 
temperatures~\cite{Ale12D,Ale14A}, although careful volume studies may prove
difficult in that regime. However, such infrared depletion has not been 
seen yet in the 12--flavor case, not even on small volumes. Thus, 
the possibility of vanishing $m_c$, with massless N$_f$=12 ending up chirally 
broken, should certainly be viewed as open.


\begin{theacknowledgments}
  We are indebted to Anna Hasenfratz and David Schaich for sharing their staggered fermion
  ensembles. Ivan Horv\'ath acknowledges the support from the Department of Anesthesiology 
  at the University of Kentucky. Andrei Alexandru is supported by U.S. National Science 
  Foundation under CAREER grant PHY-1151648.
\end{theacknowledgments}

\end{document}